%% file: Paper_ML_SLR_Updates.tex
\documentclass[conference]{IEEEtran}
\usepackage{array}
\usepackage{booktabs} 
\usepackage{multirow}
\usepackage{lipsum}
\usepackage{algorithm}
\usepackage{color, soul}
\usepackage{amsmath}
\usepackage{graphicx}
\usepackage{mdframed}
\usepackage{url}
\usepackage{cite}
\usepackage[noend]{algpseudocode}
\usepackage[utf8]{inputenc}
\usepackage{comment}
\usepackage{hyperref}

\newcolumntype{P}[1]{>{\centering\arraybackslash}m{#1}}

\ifCLASSINFOpdf
\else
\fi
\begin{document}
\title{Can Machine Learning Support the Selection of Studies for Systematic Literature Review Updates?}

\author{

\IEEEauthorblockN{Marcelo Costalonga}
\IEEEauthorblockA{\textit{PUC-Rio} \\
Rio de Janeiro, Brazil \\
mcardoso@inf.puc-rio.br}
\and
\IEEEauthorblockN{Bianca Minetto Napole\~ao}
\IEEEauthorblockA{\textit{Université du Québec à Chicoutimi} \\
Chicoutimi, Canada \\
bianca.minetto-napoleao1@uqac.ca}
\and
\IEEEauthorblockN{Maria Teresa Baldassarre}
\IEEEauthorblockA{\textit{University of Bari} \\
Bari, Italy \\
mariateresa.baldassarre@uniba.it}
\and
\IEEEauthorblockN{Katia Romero Felizardo}
\IEEEauthorblockA{\textit{Universidade Tecnológica Federal do Paraná} \\
Cornélio Procópio, Brazil \\
katiascannavino@utfpr.edu.br}
\and
\IEEEauthorblockN{Igor Steinmacher}
\IEEEauthorblockA{\textit{Northern Arizona University} \\
Flagstaff, USA \\
igor.steinmacher@nau.edu}
\and
\IEEEauthorblockN{Marcos Kalinowski}
\IEEEauthorblockA{\textit{PUC-Rio} \\
Rio de Janeiro, Brazil \\
kalinowski@inf.puc-rio.br}
}




\maketitle






\begin{abstract}
[Background] Systematic literature reviews (SLRs) are essential for synthesizing evidence in Software Engineering (SE), but keeping them up-to-date requires substantial effort. Study selection, one of the most labor-intensive steps, involves reviewing numerous studies and requires multiple reviewers to minimize bias and avoid loss of evidence.
[Objective] This study aims to evaluate if Machine Learning (ML) text classification models can support reviewers in the study selection for SLR updates.
[Method] We reproduce the study selection of an SLR update performed by three SE researchers. We trained two supervised ML models (Random Forest and Support Vector Machines) with different configurations using data from the original SLR. We calculated the study selection effectiveness of the ML models for the SLR update in terms of precision, recall, and F-measure. We also compared the performance of human-ML pairs with human-only pairs when selecting studies.
[Results] The ML models achieved a modest F-score of 0.33, which is insufficient for reliable automation. However, we found that such models can reduce the study selection effort by 33.9\% without loss of evidence (keeping a 100\% recall). Our analysis also showed that the initial screening by pairs of human reviewers produces results that are much better aligned with the final SLR update result. [Conclusion] Based on our results, we conclude that although ML models can help reduce the effort involved in SLR updates, achieving rigorous and reliable outcomes still requires the expertise of experienced human reviewers for the initial screening phase.

\end{abstract}



\input{sections/01-introduction.tex}
\input{sections/02-background.tex}

\input{sections/03-goals-and-rqs.tex}
\input{sections/04-methodology.tex}
\input{sections/05-results.tex}
\input{sections/06-discussion.tex}
\input{sections/07-threats.tex}
\input{sections/08-conclusions.tex}

\section*{Acknowledgment}

We express our gratitude to CNPq (Grant 312275/2023-4), FAPERJ (Grant E-26/204.256/2024), and Stone Co. for their generous support.

\bibliographystyle{IEEEtranS}
\bibliography{bibTex/sigproc} 

\end{document}

%% file: sections/01-introduction.tex
\section{Introduction}
\label{sec:introduction}

In the context of Evidence-Based Software Engineering (EBSE), Systematic Literature Reviews (SLR) are the main instrument to identify, synthesize, and summarize current evidence on a research topic or phenomenon of interest \cite{Kitchenham15}. Since the introduction of SLR in the Software Engineering (SE) field in 2004 \cite{Kitchenham04}, especially over the recent years, the number of SLRs has increased substantially \cite{Mendes2020,Napoleao2021S}. As stated by Mendes \textit{et al.} \cite{Mendes2020}, several SLRs in SE need to be updated. Only 20 SLRs were updated between 2006 and 2018 in a scenario of over 400 published SLRs in SE \cite{Mendes2020}. 

Outdated SLRs could lead researchers to make obsolete decisions or conclusions about a research topic \cite{Watanabe20}. Despite several initiatives in SE to keep SLRs updated (e.g., processes \cite{Dieste08a, Mendes2020}; guidelines \cite{Wohlin2020, felizardo16}; and experience reports \cite{Garces17, Felizardo20}) there is a lack of investigation on automation tools to support the SLR updates.

Performing an SLR update demands significant effort and time for reasons such as (i) the rapid growth of available evidence \cite{Zhang18, Stol15}, which hampers and slows down the identification of relevant sources; and (ii) the lack of detailed protocol documentation and data availability \cite{Ampatzoglou2019, Zhou2015}, which makes the SLR update process even more difficult, especially when different researchers carry out the update, as most of the tacit knowledge from the original SLR is lost \cite{Felizardo20, Fabbri13}. These reasons directly impact the selection of new studies during SLR updates since they are crucial to determining whether or not new evidence should be taken into account.


Machine learning (ML) has been effectively used for classification tasks in several domains.
Considering the potential benefits of exploring Machine Learning (ML) algorithms in the SLR and SLR updates context \cite{Napoleao2021, Watanabe20}, we present an empirical investigation with the \textbf{goal} of evaluating the adoption of ML models to support the selection of studies for SLR updates with respect to three perspectives: (i) effectiveness of the ML models in selecting relevant studies; (ii) effort reduction in terms of numbers of studies to be analyzed; and (iii) support in assisting reviewers in accurately selecting studies, when compared to the assistance of an additional human reviewer.

To achieve our study goal, we investigate the potential of two supervised ML models, Support Vector Machines (SVM) and Random Forest (RF). These models were chosen considering they are among the five most used text classifiers \cite{pintas2021feature} and performed better than the other three (Naive Bayes, k-Nearest Neighbors, and Decision Trees) in our initial tests. We compare their results with a rigorous manual selection process of an SLR update conducted by three experienced SE researchers. During this process, initially, each human reviewer assessed the papers based on title, abstract, and keywords using the following scale: 0 - exclude, 1 - unsure, 2 - include. Thereafter, they discussed differences and conducted the inclusion by also analyzing the full texts of the papers until reaching a consensus, producing a final curated list of included and excluded studies that we consider as our oracle. The initial assessments made by the reviewers based on the title, abstract, and keywords were naturally error-prone, similar to the ML models, which also relied on the title, abstract, and keywords for their classification. 


The main \textbf{contributions} of our study include (i) an empirical analysis of the effectiveness of ML models to support the study selection of SLR updates; (ii) the potential to reduce human effort during study selection; and (iii) an analysis on whether the ML models could assist individual reviewers in accurately selecting studies to be included.

Our findings indicated that we cannot rely on ML models for study selection in SLR updates, as their current effectiveness (f-score of 0.33) is clearly insufficient. However, there is potential for reducing the study selection effort for SLR updates by about 34\% without loss of evidence (recall 100\%), discarding studies with a low probability of being included according to the ML models. Furthermore, we found that pairing human reviewers with ML models when selecting studies is also not a good option, as their aggregated results do not improve their individual results. In fact, pairs of human reviewers performed much better than pairs of a human and an ML model.

The remainder of this study is organized as follows. Section \ref{sec:relatedwork} presents the background and related work. Section \ref{sec:researchissues} defines our goal and research questions. Section \ref{sec:methodology} describes our study design and data acquisition. Section \ref{sec:results} presents the results of each research question. Section \ref{sec:discussion} discusses our results before presenting the study's threats to validity in Section \ref{sec:threats}. Finally, conclusions are drawn (Section \ref{sec:conclusion}).

%% file: sections/02-background.tex
\section{Background and Related Work}
\label{sec:relatedwork}

An SLR update is a more recent (updated) version of an SLR \cite{Mendes2020}. For the inclusion of new and relevant evidence, one of the initial steps is to conduct the study selection activity which consists of analyzing the studies retrieved from applying the search strategy.

There have been several efforts in SE towards improvements for SLR updates (e.g. \cite{felizardo16, Garces17, Mendes2020, Wohlin2020}). However, considering the focus of our study on supporting study selection for SLR updates, we highlighted three main related works \cite{Watanabe20, Felizardo14, Napoleao2021}, described in the following.

The work of Watanabe \textit{et al.} \cite{Watanabe20} also evaluated the use of text classification (text mining combined with ML models) to support the study selection activity for SLR updates in SE. They evaluated eight SLRs from different research domains in a cross-validation procedure using Decision Tree (DT) and SVM as ML classification algorithms. The results achieved on average a \textit{F-score} of 0.92, \textit{Recall} of 0.93, and \textit{Precision} of 0.92. These results significantly outperform those of almost any expert human reviewer. However, we could not replicate their results and identified potential issues in the code. For example, feature selection was applied to the entire dataset before splitting it into training and testing sets. This means information from the SLR update may have influenced the model in ways that would not be possible before an actual SLR update. Additionally, the evaluation used time-series cross-validation without keeping the SLR update papers as a fully independent holdout set for testing purposes. While cross-validation is valuable during model development, a robust model evaluation relies on a separate test dataset to avoid data leakage, identify overfitting, and allow a fair assessment. Nevertheless, their work contributes valuable documentation of procedures and artifacts, which can help to enhance our understanding.



Felizardo \textit{et al.} \cite{Felizardo14} propose an automated alternative to support the selection of studies for SLR updates. They propose a tool called \textit{Revis}, which links new evidence with the original SLR's evidence using the K-Nearest Neighbor (KNN) Edges Connection technique. The results showed an increase in the number of studies correctly included compared to the traditional manual approach.

Napoleão \textit{et al.} \cite{minetto2024emerging} presented an automated tool prototype aimed at supporting the search and selection of studies for SLR updates using Machine Learning (ML) algorithms. Their evaluation demonstrated the feasibility of automating snowballing-based search strategies with minimal losses and that machine learning can help reduce the number of papers to be manually analyzed. Their work reinforces the potential of ML-based automation to reduce manual effort in SLR updates while highlighting the importance of conservative thresholds to minimize the risk of missing relevant studies.

More recently, Felizardo \textit{et al.}~\cite{felizardoEtAl2024} evaluated the accuracy (\textit{i.e.}, studies correctly classified) of using ChatGPT–4.0 to support SLR study selection based on the title, abstract, and keywords. Their analysis indicated that it is not advisable to outsource the selection to ChatGPT. However, they mention that it could be valuable as a support tool, aiding researchers to avoid missing evidence when properly used. However, their study did not consider SLR updates or training based on data from the original SLR.

Napoleão \textit{et al.} \cite{Napoleao2021} performed a cross-domain Systematic Mapping (SM) on existing automated support for searching and selecting studies for SLRs and SMs in SE and Medicine. The authors indicated potential ML models that can be adopted to support the study selection activities. They also indicated the most adopted methods (cross-validation during model creation and experimentation for model evaluation) and metrics (\textit{Recall}, \textit{Precision} and \textit{F-measure}) to assess text classification approaches. The choice of the ML models and the assessment metrics and methods for this study considered their findings.

%% file: sections/03-goals-and-rqs.tex
\section{Goal and Research Questions}
\label{sec:researchissues}

The goal of this study is to evaluate the adoption of ML models to support the selection of studies for SLR updates. We translated our goal into three different Research Questions (RQs).

    \textbf{RQ1:} \textit{How effective are ML models in selecting studies for SLR updates?}

    To answer this research question, we represent the effectiveness of the ML models in supporting study selection using metrics such as \textit{Recall}, \textit{Precision} and \textit{F-measure} \cite{Napoleao2021, Watanabe20}. Our ML automated analysis considers only the title, abstract, and keywords of the studies. Our ML models were trained with data from the original SLR and asked to select studies for the SLR update. The results were compared with the final results of the included and excluded studies (according to the consensus discussion of the three experienced SE researchers) for the SLR update.
    

    \textbf{RQ2:} \textit{How much effort can ML models reduce during the study selection activity of SLR updates?}

    For this research question, we calculate the effort reduction by the relation of the number of studies that need to have their title, abstract, and keywords manually analyzed without the support of ML models versus the number of studies to be analyzed after discarding studies that would have a low probability of being included according to the ML model. \textit{I.e.}, we analyzed the percentage of studies that could be safely discarded based on their inclusion probability while keeping a 100\% recall of the included studies. 
   
    \textbf{RQ3:} \textit{How does the support of ML in the selection of studies compare to the support of an additional human reviewer?}

    In this research question, we assessed how a pair of a human and an ``ML model reviewer" would compare against pairs of human reviewers in determining the list of studies to be included. Therefore, to improve the chances of providing good support, we used the ML model with the highest \textit{F-score}. In the initial screening, each of the three SE researchers had assessed each paper on a scale from zero to two (0 - exclude, 1 - unsure, 2 - include). We adjusted the outcome of the ML model (probabilities for inclusion) to that same scale of integers. Thereafter, we calculated the aggregated outcome for the list of papers for each possible pair of reviewers using the average score of the pair members. By using the average, we fairly hypothesize that, when working in pairs, each member of the pair would equally influence inclusion or exclusion. Finally, we compared the aggregated outcome of each pair with the final results using the Euclidean distance to understand how far each pair was from the oracle.




%% file: sections/04-methodology.tex
\section {Study Design}
\label{sec:methodology}




In this section, we present the study design. To evaluate our research questions, we performed a small-scale evaluation \cite{Wohlin2022cs}, following two main steps: (i) Data Collection and (ii) Design \& Execution. We describe the data collection to train and test our ML models in Section \ref{subsec:data}. In Section \ref{subsec:studydesing}, we detail how we designed and executed our strategy to train and configure the ML models. 

\subsection{Data Collection}
\label{subsec:data}

\begin{figure*} [ht]
    \centering
    \includegraphics[width=380pt]{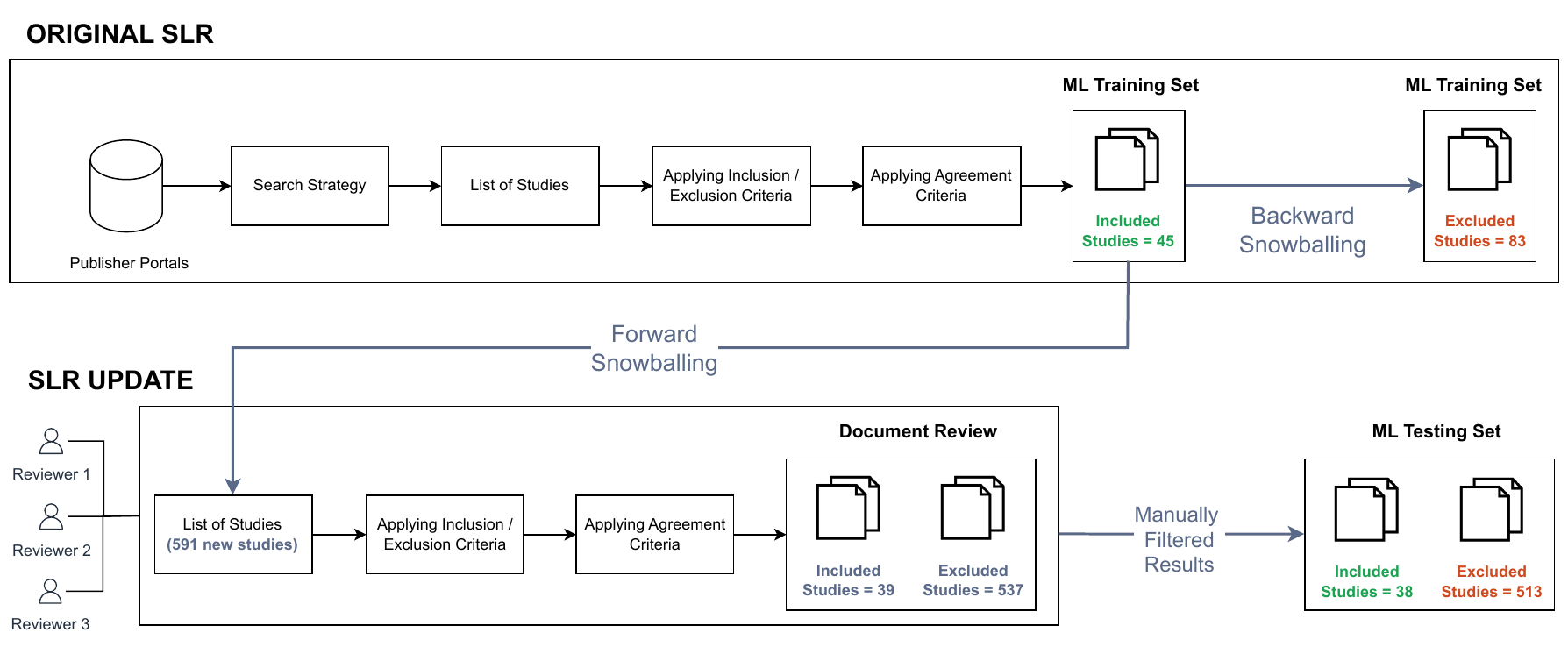}
    \caption{Data collection process}
    \label{fig:fig-data-selection}
\end{figure*}

We used an ongoing SLR update conducted by the same first three authors of this replication \cite{Wohlin2022} as the instrument of our study. We chose this ongoing SLR update since the inclusion and exclusion of new studies were rigorously conducted based on individual assessments and the consensus of three experienced SLR researchers. First, each researcher screened all the papers, analyzing the title, abstract, and keywords and registering his individual assessment (\textit{i.e.}, we had three assessments for each paper). Then, the full texts of the studies were analyzed, and discussions were held to reach a final consensus on the list of included and excluded papers. Hence, we had the results of the initial screening by each researcher and also the final list of papers to be included and excluded.

We had access to all the studies the team analyzed during the SLR update (.bib files): a total of 591 papers were analyzed for the SLR update, of which 39 were included, and 552 were excluded. We filtered the studies to consider only the studies in English and containing an abstract. In the end, we used 551 studies in our testing set, of which 38 were included by the team assessment and 513 were excluded. We used these studies to form the testing set for our ML models. 

To train our ML models, we used a training set with 128 studies, of which 45 were included and 83 were excluded. The 45 studies used to train our models with what should be included were the same studies included in the original SLR \cite{Wohlin2022}. Since we did not have access to the list of excluded studies during the study selection phase of the original SLR, we performed a backward snowballing \cite{Wohlin14} on the original references to obtain the 83 studies used to train our models with what should be excluded. Figure \ref{fig:fig-data-selection} summarizes this process. The bib files for the included and excluded studies of the training and testing sets are available in our open science repository~\cite{zenodoOpenScience}.

\subsection{Design \& Execution}
\label{subsec:studydesing}

\begin{figure*} [ht]
    \centering
    \includegraphics[width=380pt]{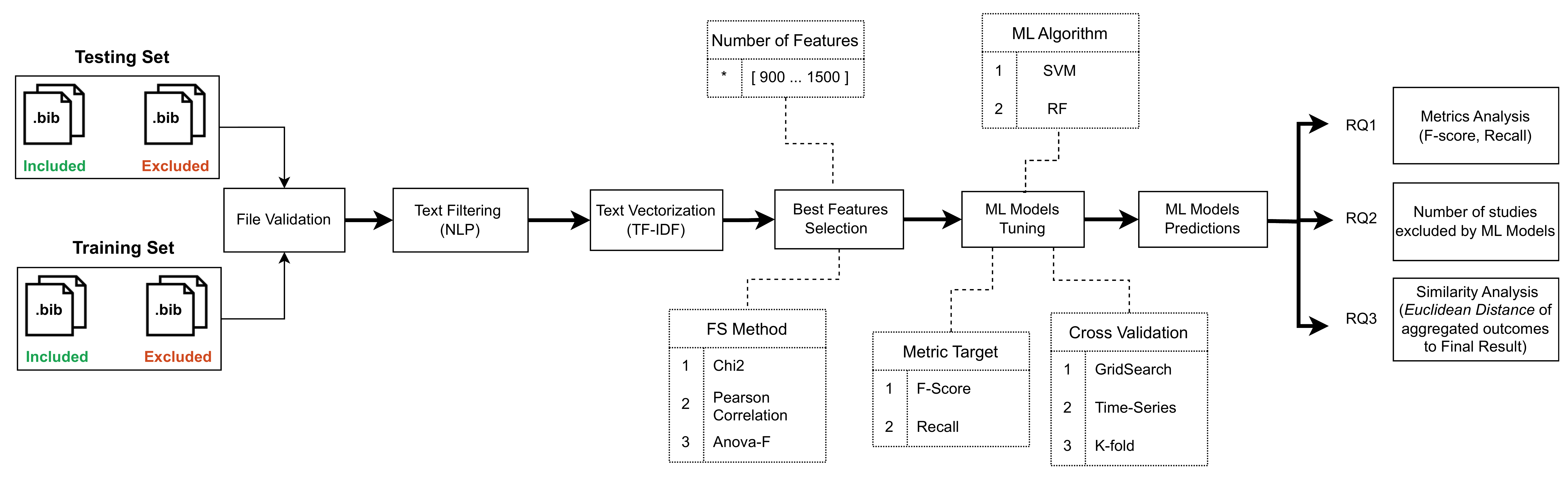}
    \caption{Study design pipeline}
    \label{fig:fig-study-design}
\end{figure*}

We developed a pipeline with the following steps to automate the study selection process of an SLR update by using ML and answering our research questions. Our pipeline is illustrated in Figure \ref{fig:fig-study-design}. 

In summary, our pipeline processes a set of .bib files containing the list of studies to train the ML models and the list of studies to be analyzed. After its execution, it returns a report file in .xlsx format with the ML model predictions, informing which studies should be included and excluded, as well as metrics about the ML model predictions and the configuration used.

The pipeline receives four different .bib files as input, one containing the list of studies that should be excluded and one containing the list of studies that should be included for each set (training and testing). In case there are any errors in the input files, the pipeline will stop its execution and will inform which entry was associated with each error as well as the type of error. 

As shown in Figure~\ref{fig:fig-study-design}, we first validated the .bib files of our testing and training sets to ensure the completeness of the set, avoiding duplicated entries or missing keys. Each study entry must have a title, the year of publication, an abstract text, and an author list. 
Secondly, we applied text filtering techniques with Natural Language Processing (NLP) \cite{NLTK}, such as Lemmatization and Tokenization, to remove irrelevant characters. Thirdly, we applied Text Vectorization on the filtered texts using  Term-frequency/Inverse-Document-Frequency (TF/IDF), a technique that transforms text data into a numerical matrix of features. Fourthly, we used statistical methods to compute and select the most relevant features. In the fifth step, we trained and tuned our ML models using our training set. Finally, in the last step, we used our ML models to predict which studies of our testing set should be included and excluded and compared the results with the final list of included and excluded studies.

Additionally, an optional .env file can be passed as input to our pipeline; this file allows some steps in our pipeline to use a specific configuration, such as choosing the configuration of the Feature Selection (FS) method to compute the features, as well as the number of features to be selected in step four, and choosing the configuration for the ML models regarding which algorithm to be used, or which metric should be targeted when tuning the model as well as the type of cross-validation to be performed, in step five. All parameters that can be configured are also shown in Figure~\ref{fig:fig-study-design}.

Building on the work by Napoleão \textit{et al.} \cite{Napoleao2021}, which highlighted Support Vector Machines (SVM) as one of the most commonly used machine learning classifiers for assisting study selection in systematic literature reviews (SLRs), we chose to evaluate SVM in our study. Additionally, inspired by the findings of Pintas \textit{et al.} \cite{pintas2021feature}, who analyzed the most widely adopted ML classifiers and feature selection techniques for text classification, identifying SVM, Naive Bayes, k-Nearest Neighbors, Decision Trees, and Random Forest (RF) as the top five classifiers, we conducted initial tests with these classifiers. Our preliminary results showed that SVM and RF outperformed the other classifiers. Consequently, we focused our evaluation on SVM and RF.


We experimented with multiple pipeline configurations and evaluated different configurations for FS and training and tuning of our ML classifiers. During step four, to compute the best features, we tested different statistical methods such as Chi-squared (Chi2) \cite{Chi2}, Pearson Correlation \cite{pearson_r} and Analysis of Variance (Anova-F) \cite{ANOVA} as well as different ranges of features. After applying Text Filtering and Text Vectorization techniques, presented in steps three and four of our pipeline, our training set comprised 23,630 features. We identified the range with the most relevant features in our training set as 900 to 1,500 features. Notably, the best results, both in terms of F-score and Recall, were consistently achieved with experiments that selected the 1200 best features.

For each evaluation, we executed the pipeline from start to finish in a clean environment using one statistical method at a time. To avoid data leakage and bias, feature selection was conducted solely based on the training set. Furthermore, we used GridSearch for parameter tuning when creating the model, which inherently includes k-fold cross-validation for measuring the most efficient parameter configuration when developing the model based on the training set. Finally, the trained model was applied to predict the inclusion or exclusion of the unseen (holdout) papers of the testing set. Hence, the hereafter reported results refer not to cross-validations conducted during model creation but to evaluating the tuned model based on the holdout testing set for which three experienced SE researchers had manually and rigorously crafted the information on inclusion and exclusion.

The complete Python code that automates our pipeline is available in our open science repository~\cite{zenodoOpenScience}.

%% file: sections/05-results.tex
\section{Results}
\label{sec:results}

To calculate the machine learning model results and address our research questions, we used two primary data sources: the final inclusion or exclusion decision for each paper in the SLR update (serving as the oracle) to answer RQ1 and RQ2, and the initial screening results from each of the three researchers to answer RQ3. Table \ref{tab:slr-update-results-sample} provides an overview of the data provided by the assessment team for 551 papers (38 included and 513 excluded). In this table, a ``Final Result" value of 1 indicates inclusion, while 0 indicates exclusion. Additionally, each reviewer recorded their initial screening assessment on a scale from zero to two (0 - exclude, 1 - unsure, 2 - include).

\begin{table}[!ht]
\scriptsize
\caption{Overview of the data provided by the assessment team.}
\begin{center}
\begin{tabular}{ m{1.8cm} m{1.5cm} m{1cm} m{1cm} m{1cm} } 
 \hline 
  \textbf{Study} & \textbf{Final Result} & \textbf{R1} & \textbf{R2} & \textbf{R3} \\
 First Study & 1 & 2 & 1 & 2 \\ 
 Second Study & 0 & 2 & 0 & 0 \\ 
 Third Study & 1 & 2 & 1 & 0 \\ 
 Fourth Study & 0 & 0 & 2 & 1 \\ 
 Fifth Study & 0 & 0 & 0 & 0 \\
 \end{tabular}
\end{center}
 \label{tab:slr-update-results-sample}
\end{table}

The complete information of our results for the best configurations we found for RQ1 and RQ2, as well as all the other test executions, are available in our open science repository~\cite{zenodoOpenScience}. 

\subsection{RQ1: \textit{How effective are ML models in selecting studies for SLR updates?}}
\label{results:RQ1}

\begin{figure}[ht]
    \centering
    \includegraphics[width=6cm]{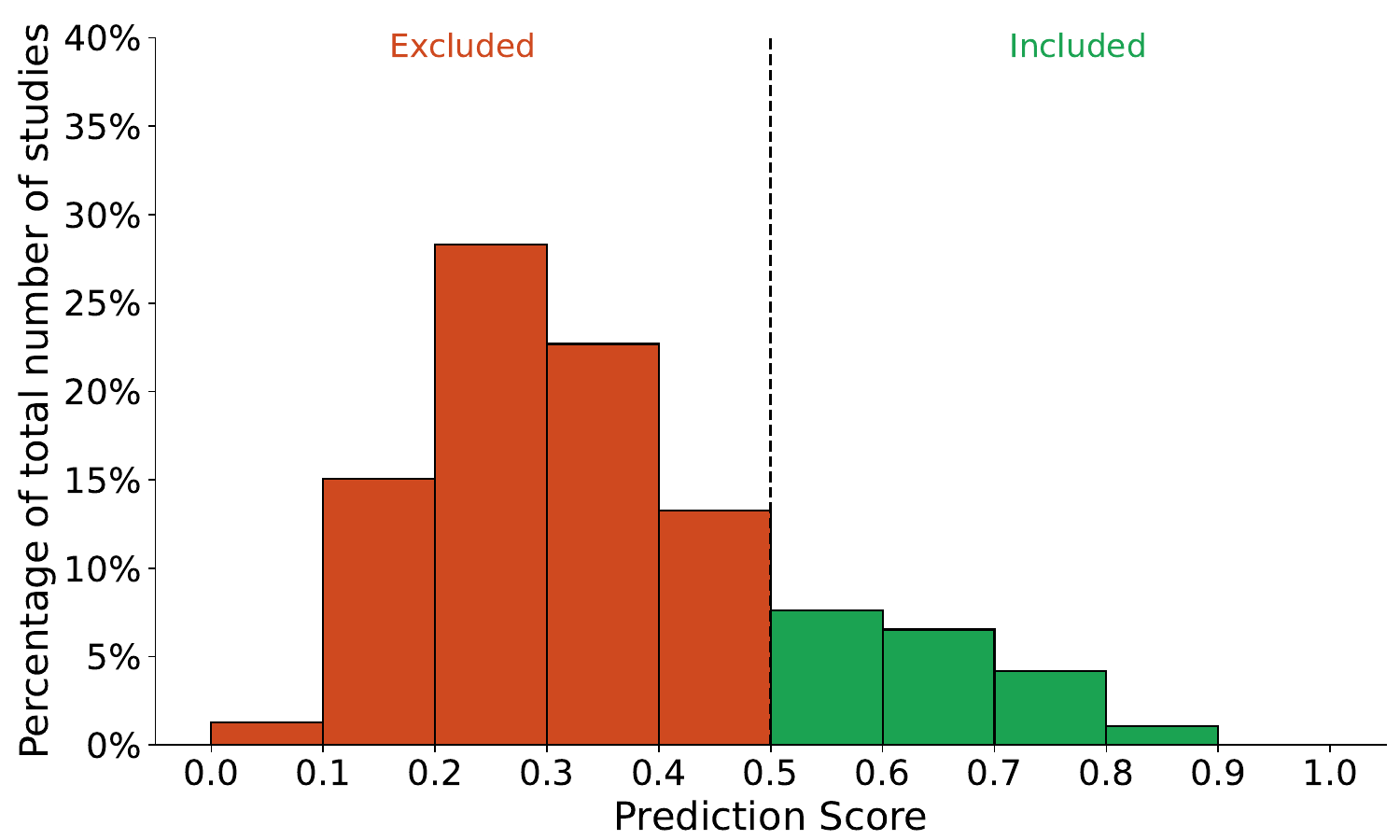}
    \caption{RF inclusion probability distribution}
    \label{fig:fig-rf-distribution}
\end{figure}

To answer this question, during the ML models tuning step, we trained our classifiers with GridSearch, focusing on maximizing the F-score. We got the best result using RF with a Precision of 0.22, Recall of 0.63, and F-score of 0.33 using the Anova-F statistical method, with 1,200 features. We used a default inclusion probability threshold of 0.5 to consider which studies should be included and excluded by our ML models. Figure \ref{fig:fig-rf-distribution} illustrates the distribution of the ML predictions' inclusion probabilities made by RF with this configuration. The parameters tested and selected by GridSearch, as well as the predictions made by our ML model using these parameters, can be found in our online repository.

Table \ref{tab:rq1-threshold-analysis} shows that by increasing the inclusion probability threshold, our RF model was able to achieve more accurate results with an F-score of 0.41 at the cost of increasing the number of FN (\textit{i.e.}, losing evidence).

\begin{table}[ht]
\centering
\scriptsize
\caption{Tradeoff between F-score and number of FN.}
\label{tab:rq1-threshold-analysis}
\begin{tabular}{|c|c|c|c|c|c|}
\hline
\textbf{Threshold(\%)} & \textbf{F-score} & \textbf{TN} & \textbf{TP} & \textbf{FN} & \textbf{FP} \\
\hline
0.50\% & 0.33 & 430 & 24 & 14 & 83 \\ 
0.60\% & 0.38 & 468 & 20 & 18 & 45 \\  
0.65\% & 0.39 & 485 & 16 & 22 & 28 \\  
0.70\% & 0.41 & 498 & 14 & 24 & 15 \\ 
\hline
\end{tabular}
\end{table}

\subsection{RQ2: \textit{How much effort can ML models reduce during the study selection activity of SLR updates?}}
\label{results:RQ2}




To answer this question, we tuned the ML models to maximize the Recall. Since the purpose of this question was to evaluate how much human effort could be reduced by using ML models during the selection of studies, we wanted to mitigate the chances of false negatives (FN) so the reviewers could simply ignore the studies excluded by the ML model without worrying about loss of evidence.

Our best result was obtained by using the SVM algorithm with a Precision of 0.10, Recall of 1.0, and F-score of 0.19 using the Pearson Correlation statistical method, with 1,200 features. We used a default inclusion probability threshold of 0.5 to consider which studies should be included and excluded by our ML models. The first line of Table \ref{tab:effort_reduction} shows these results. It is possible to observe that while maximizing the Recall to 100\%, SVM could exclude 187 studies, which represents 33.9\% of the total amount of studies in our testing set. The parameters tested and selected by GridSearch for this configuration and a table with all the predictions made by our ML model for each study analyzed to answer RQ2 are available online.  

\begin{table}[ht]
\centering
\scriptsize
\caption{Tradeoff between effort reduction and number of FN.}
\label{tab:effort_reduction}
\begin{tabular}{|c|c|c|c|c|c|c|}
\hline
\textbf{Threshold(\%)} & \textbf{RECALL (\%)} & \textbf{TN} & \textbf{TP} & \textbf{FN} & \textbf{FP} & \textbf{Reduced (\%)} \\
\hline
0.50\% & 100.00\% & 187 & 38 & 0 & 326 & 33.9\% \\ 
0.75\% & 97.37\% & 265 & 37 & 1 & 248 & 48.3\% \\  
0.80\% & 94.74\% & 283 & 36 & 2 & 267 & 51.7\% \\  
0.85\% & 89.49\% & 307 & 34 & 4 & 206 & 56.4\% \\ 
\hline
\end{tabular}
\end{table}


It is noteworthy that by increasing the inclusion probability threshold (\textit{i.e.}, including only studies for which the ML model considered a higher inclusion probability), we can reduce the human effort even more at the risk of having more FN results and losing evidence. It is possible to observe, for instance, that increasing the threshold to 75\% results in only one FN while the number of TN increases by 87. \textit{I.e.}, one study that should be included would be lost, but 87 less would have to be analyzed. Notably, this particular FN was a complicated case in which the initial screening by the human reviewers also had a lot of disparity. Reviewer R1 voted 2 (include), R2 voted 1 (unsure), and R3 voted 0 (exclude). 



\subsection{RQ3: \textit{How does the support of ML in the selection of studies compare to the support of an additional human reviewer?}}
\label{results:RQ3}

In this research question, we explored how well a human and an ``ML model reviewer" pair would perform in selecting studies for inclusion compared to pairs of human reviewers. To maximize support, we used the ML model with the highest \textit{F-score} (the same RF model used to answer RQ1). After all, a higher \textit{F-score} indicates that this model made more realistic predictions. 

\begin{figure}[ht]
    \centering
    \includegraphics[width=6cm]{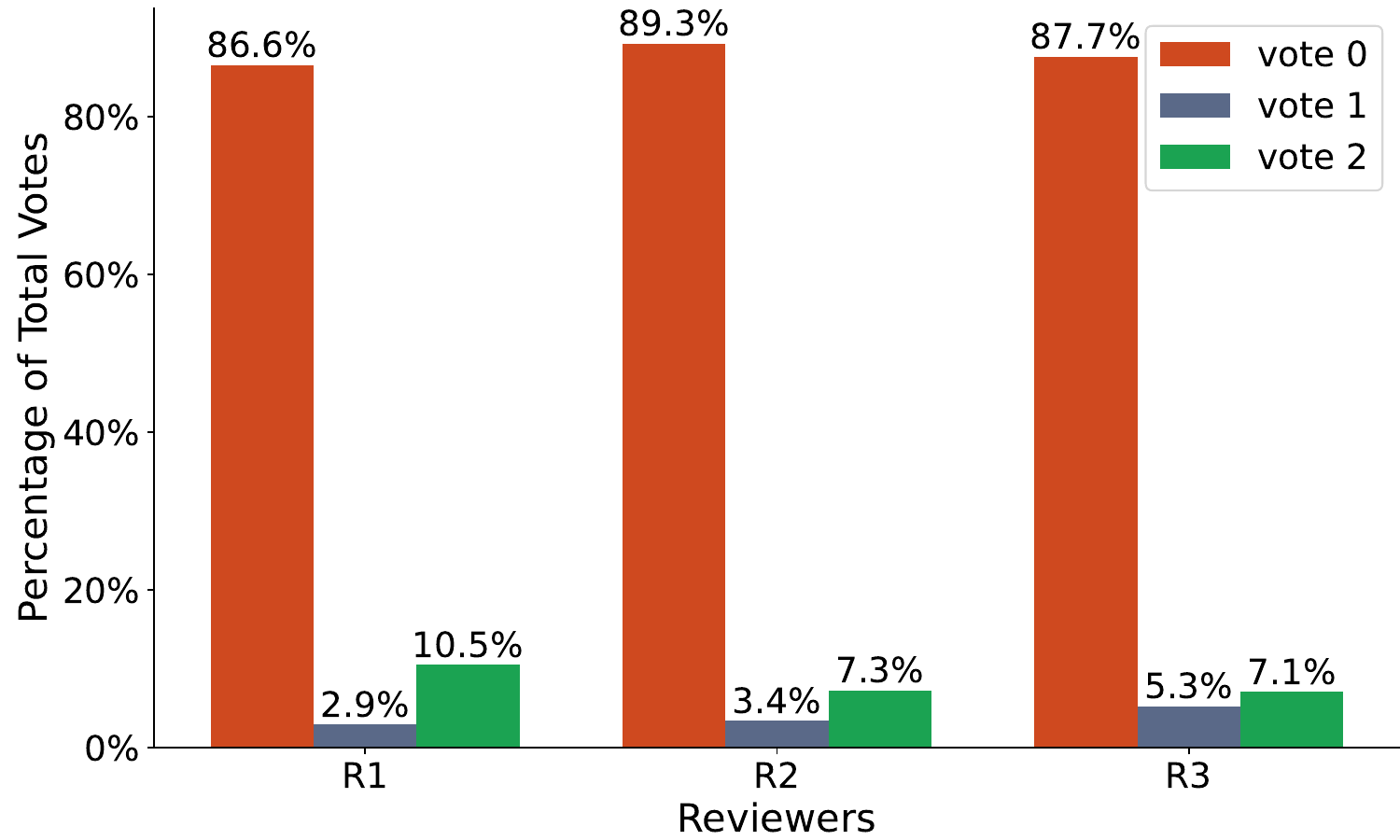}
    \caption{Reviewers assessment distribution.}
    \label{fig:fig-reviewers-votes-distribution}
\end{figure}

To prepare the data, we mapped the RF model’s inclusion probabilities to match the scale used by reviewers in the initial screening (0 - exclude, 1 - unsure, 2 - include). Since our dataset was unbalanced, as shown by the reviewers' assessment distributions in Figure~\ref{fig:fig-reviewers-votes-distribution}, we adjusted thresholds to achieve a similar distribution, mapping probabilities as follows:
\begin{itemize}
    \item 0.00 to 0.50: 0 - exclude
    \item 0.51 to 0.60: 1 - unsure
    \item 0.61 to 1.00: 2 - include
\end{itemize}

\begin{figure}[ht]
    \centering
    \includegraphics[width=6cm]{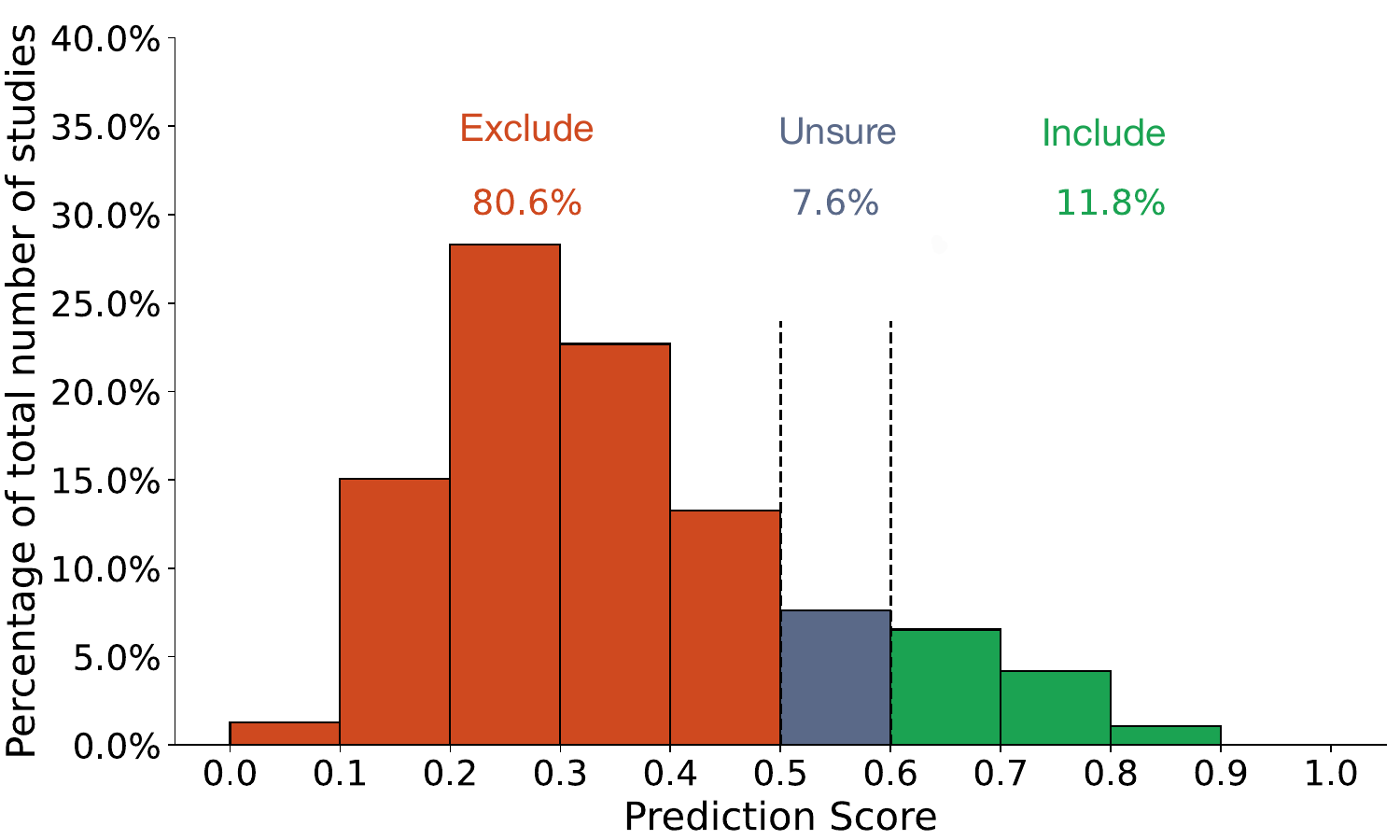}
    \caption{Predictions Distribution considering "uncertain" range.}
    \label{fig:fig-rf-normalized-distribution}
\end{figure}

Figure~\ref{fig:fig-rf-normalized-distribution} shows the RF model's probability distribution with these thresholds applied. 

By mapping the RF model’s outcomes to the same scale used by the three human reviewers, we were able to treat the model as an additional reviewer in our analyses. \textit{I.e.}, we now had initial screenings in similar scales for R1, R2, R3, and RF. The final step of data preparation involved aligning the Final Result (FR) with the same scale, assigning a value of 2 to included studies and 0 to excluded studies. 

After data preparation, our analysis involved using the Euclidean Distance (ED) to evaluate how close different combinations of reviewers were to the FR (oracle). We evaluated the ED to the FR for single reviewers, pairs, and groups of three reviewers, as follows:
\begin{itemize}
    \item Similarity between single answers and FR: 
    \[
    \textit{ED}\left(i, \text{FR}\right) \text{ where } i \in \{\text{R1}, \text{R2}, \text{R3}, \text{RF}\}
    \]
    \item Similarity between pairs and FR:
    \[
    \textit{ED}\left(\frac{i + j}{2}, \text{FR}\right) \text{ where } i \neq j \text{ and } i, j \in \{\text{R1}, \text{R2}, \text{R3}, \text{RF}\}
    \]
    \item Similarity between groups and FR:
    \begin{multline}
    \textit{ED}\left(\frac{i + j + k}{3}, \text{FR}\right) \\ 
    \text{ where } i \neq j \neq k \text{ and } i, j, k \in \{\text{R1}, \text{R2}, \text{R3}, \text{RF}\}
    \end{multline}
\end{itemize}

Table~\ref{tab:similarity-distance-FR-table} shows the ED measured in each case. As we can see, the smallest distance compared to the FR in each case was given by: R2 with ED = 9.95, pair (R2, R3) with ED = 8.86, and Group (R1, R2, R3) with ED = 8.17. It is possible to observe that (i) individual human reviewers outperformed the RF model, (ii) pairs of human reviewers alone outperformed pairs of a human reviewer and the RF ML model, and (iii) replacing any human reviewer of the assessment team with the RF model would lead to poorer results.

\begin{table}[!ht]
\centering
\scriptsize
\caption{Euclidean Distance Analysis considering FR}
\label{tab:similarity-distance-FR-table}
\begin{tabular}{|c|c|}
\hline
\textbf{Comparison} & \textbf{Euclidean Distance} \\ \hline
R1 vs FR & 12.00 \\ \hline
R2 vs FR & \textbf{9.95} \\ \hline
R3 vs FR & 11.18 \\ \hline
RF vs FR & 16.67 \\ \hline
\hline
Pair (R1,R2) vs FR & 8.90 \\ \hline
Pair (R1,R3) vs FR & 9.23 \\ \hline
Pair (R2,R3) vs FR & \textbf{8.97} \\ \hline
Pair (R1,RF) vs FR & 11.58 \\ \hline
Pair (R2,RF) vs FR & 11.48 \\ \hline
Pair (R3,RF) vs FR & 11.80 \\ \hline
\hline
Group (R1,R2,R3) vs FR & \textbf{8.25}\\ \hline
Group (RF,R2,R3) vs FR & 10.02 \\ \hline
Group (R1,RF,R3) vs FR & 9.93 \\ \hline
Group (R1,R2,RF) vs FR & 9.77 \\ \hline
\end{tabular}
\end{table}

%% file: sections/06-discussion.tex
\section{Discussion}
\label{sec:discussion}

The results provide valuable insights into the limitations of machine learning (ML) models to support systematic literature review (SLR) updates. In this discussion, we interpret these results in light of the research questions, contextualize their implications, and outline the trade-offs associated with applying ML models in this domain.

\subsection{Effectiveness of ML Models for SLR Study Selection (RQ1)}

The results for RQ1 indicate that our best-performing model, Random Forest (RF), achieved a modest balance between precision and recall with an F-score of 0.33 at the default threshold of 0.5. This result suggests that while the ML model was able to identify some relevant studies, its overall ability to precisely distinguish between relevant and irrelevant studies was limited. Adjusting the threshold improved the F-score to 0.41, highlighting the sensitivity of the model’s performance to the chosen threshold. However, this improvement came at the cost of increasing false negatives (FNs), potentially missing valuable studies. We interpret the RF model’s performance as indicating that ML may assist in informally identifying a subset of relevant studies but is not yet reliable for the selection of studies for SLR updates.

\subsection{Effort Reduction through ML Models (RQ2)}

In answering RQ2, we focused on maximizing recall to avoid FNs. In our investigations, the SVM model was more suitable for focusing on achieving a high recall and demonstrating some potential for reducing human screening efforts. Results demonstrated that with a recall of 100\%, the SVM model could exclude 33.9\% of studies from the review process without missing any relevant studies. This reduction represents a significant decrease in the manual workload, suggesting ML’s potential to assist researchers with the initial screening stage. However, to achieve this high recall, the model produced a high rate of false positives (FPs), still requiring significant human review effort to discard many non-relevant studies.

As shown in Table \ref{tab:effort_reduction}, gradually increasing the inclusion probability threshold reduced the number of FPs at the cost of a minor drop in recall. For instance, at a threshold of 0.75, the model achieved a recall of 97.37\%, with a reduction of 48.3\% in the number of studies needing review. We interpret this result as indicating that, while ML can reduce screening efforts, care must be taken when applying thresholds to avoid introducing a risk of overlooking critical studies.

\subsection{Supporting Human Reviewers (RQ3)}

For RQ3, we evaluated the support ML could provide compared to that of an additional human reviewer. When we treated the RF model as an additional reviewer and calculated Euclidean Distance (ED) to assess alignment with the final inclusion decision, individual human reviewers outperformed the RF model. Furthermore, pairs of human reviewers clearly outperformed human-ML pairs, suggesting that human-only review teams achieve more accurate results.

This finding reinforces the challenges ML models face in fully replicating the nuanced judgment of human reviewers. Hence, ML can not replace additional human reviewers, and ML assistance is not a valid argument for quality in the selection process. Pairs of human reviewers are still highly recommended for selecting studies in SLR updates.

%% file: sections/07-threats.tex
\section{Threats to Validity}
\label{sec:threats}

In the following, we enumerate the main threats to the validity of our study, using the categories suggested by~\cite{Runeson12}.

\textbf{Construct Validity.}
Our evaluation results might have been affected by the choice of the SLR update and of the ML algorithms. Regarding the chosen SLR update, it is very difficult to get access to details such as individual assessments by reviewers during the initial screening process, which we needed for our analyses. Our SLR update dataset had such detailed information for 551 studies and is available online~\cite{zenodoOpenScience}. Regarding the algorithms, we analyzed the most used ones for text classification~\cite{pintas2021feature} and dug deeper into the two that showed the most prominent initial evaluation results on our dataset. 

\textbf{Internal Validity.}
Our training dataset comprised only studies included in the SLR replication \cite{Wohlin2022} (training included) and those obtained through backward snowballing (training excluded). We deliberately excluded studies not in English or those categorized as Ph.D. dissertations or book chapters from our testing set, the same criteria adopted by the SLR. A potential threat to internal validity that could have favored human reviewers is that, during the manual initial screening process, while this was not part of the procedure, the human reviewers could have ended up reading other sections of the studies besides the title, abstracts, and keywords.

\textbf{External Validity.} The dataset used in our analysis might not represent the diversity of SLR updates in SE. However, we did not find other SLRs with available data on the individual assessments applied during the initial screening. Replicating the investigations on other SLR updates to strengthen external validity would require significant effort for which we would have to involve the wider community. While not claiming external validity, we believe that sharing our initial evaluation results can already provide some valuable insights.

\textbf{Reliability.} The data used in our evaluation, including the individual initial screening assessments and the final list of papers to be included in the SLR update, was generated by the same (first three) authors who performed the SLR replication~\cite{Wohlin2022}. In addition, to improve the reliability of our results, our ML models and the evaluation datasets are openly available and auditable. 

%% file: sections/08-conclusions.tex
\section{Conclusion}
\label{sec:conclusion}

This study investigated the application of supervised ML models as a supporting tool for researchers during study selection in SLR updates. Therefore, we developed a supervised ML pipeline for the study selection process. The focus was on investigating the effectiveness of ML models, the potential to reduce human effort, and the ability to provide support to individual human reviewers. We employed two ML models, Random Forest (RF) and Support Vector Machine (SVM), and assessed them on a dataset derived from a carefully manually curated SLR update. During this investigation process, our work also highlighted different configurations used for our ML models that correlate to their recall and F-score, providing results that can be useful for further exploration in this area.

Our results indicate that while ML can assist in preliminary study selection by reducing the volume of studies requiring manual review, it is not yet effective enough to automate this process or to directly assist a single human reviewer to produce more accurate selection results. Specifically, RF, our best model for study selection effectiveness, achieved a modest F-score of 0.33, with limited precision and recall, which is clearly insufficient for study selection. Meanwhile, SVM demonstrated potential in reducing effort by excluding up to 33.9\% of irrelevant studies without sacrificing recall. The comparison between human-only reviewer pairs and human-ML reviewer pairs for the initial screening showed that pairs of human reviewers produce results that are much better aligned with the final curated result of the SLR update. 

Considering our findings, we put forward that serious SLR update efforts should still rely on (at least two) experienced human researchers for the initial screening of papers to be included. Hence, this study contributes to understanding the practical limitations of ML in study selection and highlights the need for careful human involvement in this process to ensure the quality and rigor of SLR outcomes. 

Future research could focus on refining ML configurations, investigating adaptive thresholds to improve model performance in SLR update contexts, and exploring hybrid approaches (\textit{e.g.}, humans assisted by ML to reduce the overall screening effort by discarding studies with low probability of being included). We also recommend further investigating large language models (LLMs) within the SLR update context.